\documentclass[prl,superscriptaddress,unsortedaddress,twocolumn,%
  showpacs,preprintnumbers,amsmath,amssymb]{revtex4-1}
\usepackage{graphicx}
\usepackage[breaklinks, colorlinks=true, pdfstartview=FitV, linkcolor=red,%
 citecolor=blue, urlcolor=blue]{hyperref}

\newcommand{\LQCD}{\Lambda_{\text{QCD}}}
\newcommand{\piLambda}{\Lambda_\pi}
\newcommand{\MeV}{\;\text{MeV}}
\newcommand{\GeV}{\;\text{GeV}}
\newcommand{\Tc}{T_{\text{c}}}
\newcommand{\muq}{\mu_{\text{q}}}
\newcommand{\Gc}{G_{\text{c}}}
\newcommand{\Lag}{\mathcal{L}}

\newcommand{\tr}{\text{tr}}
\newcommand{\feyn}[1]{
  \setbox0=\hbox{\ensuremath{#1}}
  \hbox to\wd0{\hbox to0pt{\hbox to\wd0{\hss/\hss}\hss}\box0}}

\begin{document}
\title{Magnetic Catalysis vs Magnetic Inhibition}
\preprint{RIKEN-QHP-43}
\author{Kenji Fukushima}
\affiliation{Department of Physics, Keio University, %
             Kanagawa 223-8522, Japan}

\author{Yoshimasa Hidaka}
\affiliation{Theoretical Research Division, Nishina Center, RIKEN, %
             Wako 351-0198, Japan}

\begin{abstract}
 We discuss the fate of chiral symmetry in an extremely strong
 magnetic field $B$.  We investigate not only quark fluctuations but
 also neutral meson effects.  The former would enhance the
 chiral-symmetry breaking at finite $B$ according to the Magnetic
 Catalysis, while the latter would suppress the chiral condensate once
 $B$ exceeds the scale of the hadron structure.  Using a chiral model
 we demonstrate how neutral mesons are subject to the dimensional
 reduction and the low dimensionality favors the chiral-symmetric
 phase.  We point out that this effect, the Magnetic Inhibition, can
 be a feasible explanation for recent lattice-QCD data indicating the
 decreasing behavior of the chiral-restoration temperature with
 increasing $B$.
\end{abstract}

\maketitle


Theorists have been pursuing the answer of questions in various
extremes.  What happens if the temperature is extremely high?
According to the theory of the strong interaction, namely, quantum
chromodynamics (QCD), chiral symmetry should be restored and color
degrees of freedom should be released at the temperature $T$ of the
order of
$\LQCD\sim 200\MeV$~\cite{Hatsuda:1994pi,Rischke:2003mt,Fukushima:2011jc}.
If $T$ is raised further, the electroweak phase transition should take
place at $T\sim 100\GeV$.  Along the same spirit many theorists are
trying to clarify what could happen at extremely high baryon density.
In the QCD asymptotic regime where the perturbative calculation should
work, theoretical considerations predict the color superconducting
phases~\cite{Rischke:2003mt,Alford:2007xm,Fukushima:2010bq}.  In
particular the ground state should form the color-flavor locking
(CFL) if the quark chemical potential, $\muq$, is sufficiently larger
than the strange quark mass, while there may appear many other pairing
patterns such as the uSC phase, the dSC phase, the 2SC phase, etc in
the intermediate density region.  It is still a big theoretical
challenge to identify the correct phase structure of QCD on the whole
$\muq$-$T$ plane.

Recently the QCD phase structure in the presence of strong magnetic
field $B$ has been revisited extensively for several good
reasons~\cite{Fukushima:2011jc}:
First, QCD matter under strong $B$ is worth thinking as a realistic
situation in non-central collisions in the relativistic heavy-ion
experiment as conducted at the Relativistic Heavy-Ion Collider (RHIC)
and the Large Hadron Collider (LHC).  A simple calculation gives us an
order estimate for the produced magnetic field as $eB > \LQCD^2$ at
the RHIC energy~\cite{Kharzeev:2007jp,*Skokov:2009qp,*Deng:2012pc}.
Second, the $B$-effect is very similar to the baryon chemical
potential in a sense that gluons have no direct coupling and the QCD
equation of state is affected only through quark polarization
processes.  The Monte-Carlo simulation with $B$ is, in spite of the
similarity to $\muq$, possible without the notorious sign
problem~\cite{D'Elia:2010nq,Bali:2011qj,*Bali:2012zg}, which is a
great theoretical advantage.  Third, it is certainly intriguing to
address such a simple and well-defined question;  what is the ground
state of QCD matter when a very strong magnetic field, $B\gg\LQCD$, is
applied?

It should be tough in general to answer to such a question since the
confinement/deconfinement phenomena belong to the gluon dynamics and
$B$-effects are then indirect.  (See
Ref.~\cite{Galilo:2011nh,*Fraga:2012fs,*Fraga:2012ev} for recent
attempts.)  This difficulty is common also in the finite-$\muq$
analysis, which hinders the QCD phase diagram research
also~\cite{Schaefer:2007pw,*Fukushima:2010is}.  In other words, one
could never reach the correct QCD phase diagram on the $\muq$-$T$
plane until one can establish a machinery to reveal the $B$-effects
that should be under better theoretical
control~\cite{Mizher:2010zb,*Gatto:2010pt}.  In a related context the
new entanglement with finite $B$ and $\muq$ is also an interesting
research subject~\cite{Preis:2010cq,*Preis:2012fh}.

In contrast to the confinement sector, the properties of chiral
symmetry reside in the quark part, and thus they are directly
sensitive to the presence of $B$.  It would be, therefore, more
tractable to focus on chiral symmetry in the strong $B$ limit.  From
this point of view of the interplay between chiral symmetry and $B$,
the \textit{Magnetic Catalysis} is one of the most significant
phenomena~\cite{Klimenko:1990rh,*Klimenko:1992ch,Gusynin:1994re,%
*Gusynin:1994va,*Gusynin:1994xp,*Gusynin:1995nb,Shushpanov:1997sf}:
In chiral quark models such as the Nambu--Jona-Lasinio (NJL) model
without the confinement effect, chiral symmetry is spontaneously
broken for a sufficiently strong coupling constant, i.e.\ $G>\Gc$.
With the magnetic field, on the other hand, a non-zero value of the
chiral condensate would be an inevitable consequence from the Landau
zero-mode contribution regardless of the value of $G$.  Fermions are
always massive in the presence of $B$, hence, even though they are
massless at the Lagrangian level (i.e.\ the chiral limit).  In fact,
the (3+1) dimensional NJL model with
$\mathrm{U(1)_L}\times \mathrm{U(1)_R}$ chiral
symmetry~\cite{Gusynin:1994re} [this is the unbroken part of chiral
  symmetry with $B$ that breaks isospin symmetry explicitly] is
defined by the Lagrangian,
\begin{equation}
 \Lag = \bar{\psi} i\feyn{D}\psi + \frac{G}{2}\Bigl[
 (\bar{\psi}\psi)^2 + (\bar{\psi} i\gamma_5 \psi)^2 \Bigr] \;,
\end{equation}
where $\feyn{D}\equiv \gamma^\mu(\partial_\mu+ieA_\mu)$ with
$A^\mu=(0,-yB/2,xB/2,0)$ (in the symmetric gauge) and $B>0$.  Then it
has been established that the constituent quark mass is expressed
analogously to the gap obtained in the BCS theory;
\begin{equation}
 m^2 = \frac{eB}{\pi}\exp\biggl(-\frac{4\pi^2}{eB G}\biggr) \;,
\label{eq:quark-mass}
\end{equation}
which is non-zero for any $G$.  One can understand
Eq.~\eqref{eq:quark-mass} in such a way that $B$ is a catalyst to
induce a non-vanishing chiral condensate, that is, the Magnetic
Catalysis.  In the NJL model as an effective description of QCD at low
energy, $G$ has the energy scale comparable to $\LQCD$, i.e.\
$G\sim\LQCD^{-2}$.  This means, together with
Eq.~\eqref{eq:quark-mass}, that the $B$-induced value of $m^2$ becomes
appreciable for $eB \sim \LQCD^2$.

It is natural, as suggested by the Magnetic Catalysis, that $B$ should
enhance the chiral-symmetry breaking, so that the critical
temperature, $\Tc$, for chiral restoration should increase with
increasing $B$, which is indeed the case in all chiral-model
calculations (see Ref.~\cite{Gatto:2012sp} for a recent review).  The
latest lattice-QCD data, however, supports dropping behavior of $\Tc$
as a function of $B$ and there is no clear physical 
explanation for this.  The main goal of the present work is to propose
a new mechanism for chiral restoration at strong $B$ that could be a
feasible explanation for the lattice-QCD data.

Our idea is as follows.  Because
$\mathrm{U(1)_L}\times \mathrm{U(1)_R}$ is spontaneously broken down
to $\mathrm{U(1)_V}$, the Nambu-Goldstone (NG) boson (i.e.\ $\pi^0$)
must exist as a composite of fermions.  It does not matter whether
$\pi^0$ is a tight bound-state particle of QCD or not, but here let us
just call this NG boson $\pi^0$ in our QCD-based convention.  If $B$
is extremely strong, fermions that form $\pi^0$ are affected by $B$
and eventually their motions are restricted along the $B$-direction.
Hence, it should be conceivable that $\pi^0$ also undergoes the
dimensional reduction to the (1+1)-dimensional dynamics.  Once this
happens, the spontaneous chiral-symmetry breaking is prohibited
according to Marmin-Wagner's
theorem~\cite{Mermin:1966fe,*Coleman:1973ci}.  Such a possibility,
that we name the \textit{Magnetic Inhibition}, was considered
partially in Ref.~\cite{Gusynin:1994xp}, but only the approximated
form of the $\pi^0$ propagator was discussed there.  In this work we
will fully evaluate the $\pi^0$ propagator to address its dynamical
change in strong $B$ and formulate the above-mentioned idea.


For later convenience let us look closely at the derivation of
Eq.~\eqref{eq:quark-mass} using the quark propagator on top of the
vector potential $A^\mu$.  The quark propagator is expressed as
$S(p) = \sum_{n=0}^\infty i S_n(p)/(p_\parallel^2-m_n^2)$ with
$m_n^2\equiv m^2+2eBn$ and
$S_n(p)\equiv (\feyn{p}_\parallel + m) \bigl[P_+ A_n(p_\perp^2)
 +P_- A_{n-1}(p_\perp^2)\bigr] + \feyn{p}_\perp B_n(p_\perp^2)$.
Here we have introduced several notations:
$p_\parallel^2\equiv p_0^2 - p_z^2$, $p_\perp^2\equiv p_x^2+p_y^2$,
$A_n(p_\perp^2)\equiv 2\, e^{-2z}(-1)^n L^{(0)}_n(4z)$, and
$B_n(p_\perp^2)\equiv 4\, e^{-2z}(-1)^n L^{(1)}_{n-1}(4z)$ with
$z\equiv p_\perp^2/(2eB)$ and the projection operators;
$P_\pm \equiv (1\pm i \gamma^1\gamma^2)/2$.  The generalized Laguerre
polynomials are defined as usual by
$L_n^{(\alpha)}(x)\equiv (e^x x^{-\alpha}/n!)(d^n/dx^n)(e^{-x} x^{n+\alpha})$%
~\cite{Gusynin:1994re,Chodos:1990vv}.

The most important ingredient to study the Magnetic Catalysis is the
gap equation at the quark one-loop level,
\begin{align}
 0 &= \frac{m}{G} -
 \int\frac{d^4 p}{(2\pi)^4}
  \;\tr S(p) \notag\\
 &=  \frac{m}{G} - \frac{m}{2\pi} \cdot \frac{eB}{2\pi}
\int_{1/\Lambda^2} \frac{ds}{s}\, e^{-m^2 s}\coth(eB s) \;,
\end{align}
which is regularized in the proper-time method.

\begin{figure}
 \includegraphics[width=0.75\columnwidth]{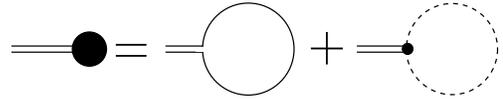}
 \caption{Schematic expression of the gap equation.  The solid curve
   represents the quark loop with the quark propagator $S(p)$ and
   the four-Fermi coupling $G$ and the dashed curve represents the
   pion loop with the pion propagator $D_\pi$ and the coupling
   $\Gamma_{\sigma\pi\pi}$.}
 \label{fig:diag}
\end{figure}

This represents the tadpole diagram in terms of quarks (see the loop
diagram with a solid line in Fig.~\ref{fig:diag}).  Obviously $m=0$ is
a solution of the above equation.  We have to make a comparison in
energy to locate the ground state that should have the lowest energy.
It is a straightforward exercise to evaluate the effective potential
by integrating the gap equation with respect to $m$, which leads to
$V(m)=m^2/(2G)+V_{\text{q}}(m)$, where the quark part is
\begin{equation}
 V_{\text{q}}(m) = \frac{eB}{8\pi^2}\int_{1/\Lambda^2}
  \frac{ds}{s^2}\; e^{-m^2 s} \coth(eBs) + \text{(const.)} \;.
\end{equation}
The onset for the spontaneous chiral-symmetry breaking in the chiral
limit is immediately located from the sign of the potential curvature,
i.e.\ the coefficient of the $m^2$-term in the potential.  In the
$B=0$ case, we can expand as
$V_{\text{q}}(m) \simeq -(\Lambda^2/8\pi^2)m^2+\text{(const.)}$, 
from which we can conclude that chiral symmetry is spontaneously
broken only for $G\Lambda^2 > 4\pi^2$.  [Note that we did not consider
  the color factor here.]  In the limit of $eB\gg\Lambda^2$, on the
other hand, only the lowest Landau level (Landau zero-mode) can
contribute to the gap equation, or, we can approximate the gap
equation as $\coth(eBs)\simeq 1$ to find
\begin{equation}
 V_{\text{q}}(m) \simeq
  \frac{eB}{8\pi^2}\biggl(\Lambda^2 - m^2
  \ln\frac{e^{1-\gamma}\Lambda^2 }{m^2} + \mathcal{O}(m^3)
  \biggr) \;.
\label{eq:V-LLL}
\end{equation}
We see that the potential curvature has a logarithmic singularity at
$m=0$ and thus the curvature can be always negative for sufficiently
small $m$, which means that the symmetric state with $m=0$ cannot be
realized.  This is how the Magnetic Catalysis works.  The extremal
point of $m^2/(2G)+V_{\text{q}}(m)$ with Eq.~\eqref{eq:V-LLL} gives a
gap equation whose solution reads
$m^2=e^{-\gamma}\Lambda^2 e^{-4\pi^2/(eB G)}$.  We note that the
difference in the overall coefficient from Eq.~\eqref{eq:quark-mass}
originates from whether $eB\gg\Lambda^2$ or not.  As long as
$eB\ll\Lambda^2$, the dynamical quark mass is characterized by $eB$,
but once $eB$ exceeds the order of $\Lambda^2$, the quark mass squared
is no longer proportional to $eB$ but suppressed by another (smaller)
scale of $\Lambda^2$.

Now let us proceed to the calculations including the pion-loop effects
under strong $B$.  Because charged pions are as massive as $eB$, we
can simply discard $\pi^\pm$ and focus only on $\pi^0$, which
justifies the usage of our simple model setting with only
$\mathrm{U(1)_L}\times\mathrm{U(1)_R}$ except for the color factor.
As long as $eB$ is small as compared to the pion size, we can treat
$\pi^0$ as a point particle as in the chiral perturbation
theory~\cite{Shushpanov:1997sf,Agasian:2008tb,*Andersen:2012zc,%
*Andersen:2012dz}.  However, $\pi^0$ is a composite particle, and it
is conceivable that the dispersion relation of $\pi^0$ should be
significantly modified by $B$.  We can concretely investigate this
by constructing $\pi^0$ dynamically in the present model.  In the
conventional random phase approximation the pion propagator is
\begin{equation}
 iD_\pi^{-1}(p) = -\frac{1}{G} + i\int\frac{d^4 k}{(2\pi)^4}\;
  \tr\bigl[ \gamma_5 S(k) \gamma_5 S(p+k) \bigr] \;.
\end{equation}
After some (tedious) calculations we can find the following
expression:
\begin{align}
 iD_\pi^{-1} &= -\hat{m}_\pi^2 + \frac{eB}{2\pi} e^{-z} \sum_{n,l=0}^\infty
  \frac{l!}{n!} z^{n-l} i \Pi_2(p_\parallel^2,m_n^2,m_l^2) \notag\\
 &\qquad\qquad\qquad \times
  \Bigl[ p_\parallel^2 F_{nl}^\parallel(z)
  - p_\perp^2 F_{nl}^\perp(z) \Bigr] \;,
\label{eq:pion-prop}
\end{align}
where we introduced new notations to indicate some combinations of the
Laguerre polynomials,
$F_{nl}^\parallel(z)\equiv[L_l^{(n-l)}(z)]^2+(n/l)[L_{l-1}^{(n-l)}(z)]^2$
and
$F_{nl}^\perp(z)\equiv (z/l)[L_{l-1}^{(n-l+1)}(z)]^2
 + (n/z)[L_l^{(n-l-1)}(z)]^2$.  Also, we defined
\begin{equation}
\Pi_2(p_\parallel^2,m_n^2,m_l^2)\equiv \int\frac{d^2 k_\parallel}{(2\pi)^2}
\frac{i}{k_\parallel^2-m_n^2}\frac{i}{(k_\parallel+p_\parallel)^2-m_l^2} \;,
\end{equation}
with $m_n^2=m^2+2eBn$ and $m_l^2=m^2+2eBl$.  The expression looks
complicated and it would be convenient to approximate it as
\begin{equation}
 iD_\pi^{-1} \approx Z_\pi^{-1}\bigl( p_\parallel^2
  - v_\perp^2 p_\perp^2 - m_\pi^2 \bigr) \;,
\label{eq:pion-prop-approx}
\end{equation}
that is motivated from an expansion valid for $p_\perp^2 < eB$ and
$p_\parallel^2 < m^2$.  The explicit computation gives
$Z_\pi^{-1}=(1/8\pi^2)[eB/m^2 + \ln(e^{-\gamma}\Lambda^2/2eB)
 -\psi(1+m^2/2eB)]$ and
$v_\perp^2=(Z_\pi/8\pi^2)\ln(e^{-\gamma}\Lambda^2/m^2)$ with $\psi(x)$
being the digamma function~\cite{Gusynin:1994re}.  We note that the
physical pion mass is given by
$m_\pi^2\equiv Z_\pi\hat{m}_\pi^2$ using the bare pion mass
$\hat{m}_\pi$ in Eq.~\eqref{eq:pion-prop}.

\begin{figure}
 \includegraphics[width=0.9\columnwidth]{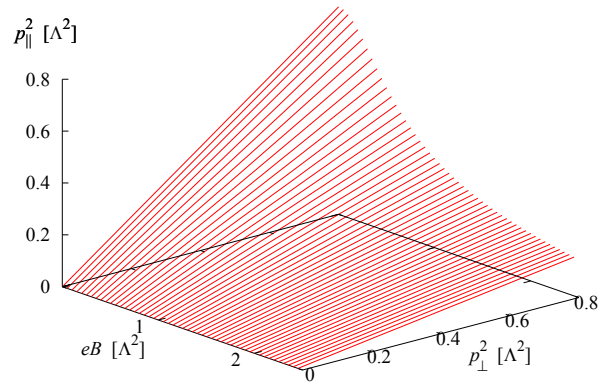}
 \caption{Zero of the pion propagator inverse~\eqref{eq:pion-prop} as
   a function of $p_\parallel^2$, $p_\perp^2$, and $eB$ in the unit of
   $\Lambda^2$.  We chose $m=0.5\Lambda$ to avoid unphysical threshold
   effects.  The slope of $p_\parallel^2$ against $p_\perp^2$
   corresponds to the transverse velocity $v_\perp^2$.  Clearly the
   expanded form~\eqref{eq:pion-prop-approx} is a good approximation
   even for $p_\perp^2 > eB$ and/or $p_\parallel^2 > m^2$.}
 \label{fig:velocity}
\end{figure}

To see how Eq.~\eqref{eq:pion-prop-approx} works, we numerically
evaluate the full propagator~\eqref{eq:pion-prop} to make a plot for
the dispersion relation in Fig.~\ref{fig:velocity}.  As long as
Eq.~\eqref{eq:pion-prop-approx} is a sensible approximation of
Eq.~\eqref{eq:pion-prop}, the zero of the pion propagator inverse
should behave like $p_\parallel^2 = v_\perp^2 p_\perp^2 + m_\pi^2$,
which is clearly confirmed in Fig.~\ref{fig:velocity}.  Furthermore,
this type of the dispersion form persists even for $p_\perp^2 > eB$
and/or $p_\parallel^2 > m^2$.  


Concerning the properties of $Z_\pi$ and $v_\perp^2$, here, the
essential point is that $Z_\pi^{-1} \sim \mathcal{O}(eB)$ and thus
$Z_\pi$ goes smaller with increasing $eB$.  This makes the transverse
velocity of $\pi^0$ behave as $v_\perp^2 \sim Z_\pi \sim 1/(eB)$ that
goes smaller accordingly.  Such vanishing behavior of $v_\perp^2$ is
nothing but the concrete realization of the dimensional reduction from
the (3+1)- to the (1+1)-dimensional dynamics.

We would make a remark about the nature of the dimensional reduction
for quarks and pions.  One might have thought, at a first glance, that
the dimensional reduction with $v_\perp^2\to 0$ in
Eq.~\eqref{eq:pion-prop-approx} seems to be a different situation from
quarks under strong $B$.  In an intuitive picture quarks are trapped
by $B$ and the transverse motion is highly restricted, so that quarks
can move only along $B$, which is how the dimensional reduction occurs
for quarks.  On the other hand, the neutral pion costs no energy to
move and thus travels freely in the transverse directions when
$v_\perp^2=0$.  Such intuitive descriptions may sound far different
but the underlying physics is common.  Actually the Landau level for
quarks is just a quantum number, and even for the Landau zero-mode for
example, the $p_\perp$-integration should be carried out, which picks
up the Landau degeneracy factor, $eB/(2\pi)$.  In the same manner the
$p_\perp$-integration for pions should count the density of states.

We shall explicitly go into the computation of the pion loop as
depicted by the dashed line in Fig.~\ref{fig:diag}.  For
$p_\parallel^2>4m^2$ the full propagator given by
Eq.~\eqref{eq:pion-prop} would suffer from the threshold effect
associated with the $\pi^0\to q\bar{q}$ decay which is unphysical due
to the lack of confinement.  We can evade this artifact by keeping the
approximate form~\eqref{eq:pion-prop-approx}, the validity of which is
checked in Fig.~\ref{fig:velocity}.  We also cut off the transverse
momentum integration in the range $p_\perp^2 \lesssim eB$, which is
reminiscent of the coefficient in Eq.~\eqref{eq:quark-mass};  the
transverse degeneracy factor should be either $eB$ or $\Lambda^2$ that
is smaller than the other.  The microscopic origin of this cutoff by
$eB$ is the $p_\perp$-dependence in $v_\perp^2$.  Not to rely on
model-dependent details, we postulate the
form~\eqref{eq:pion-prop-approx} and introduce a sharp cutoff with an
unknown parameter $\xi$ as
\begin{align}
 &\frac{1}{2}\int_{p_\perp^2<\xi eB}\frac{d^4 p}{(2\pi)^4}
  \Gamma_{\sigma\pi\pi}(p) D_\pi(p) \notag\\
 &= 2m \int_{p_\perp^2<\xi eB}\frac{d^4 p}{(2\pi)^4}\, \frac{i}
  {p_\parallel^2 - v_\perp^2 p_\perp^2 - m_\pi^2} \notag\\
 &= 2m \int_{1/\piLambda^2}^\infty ds \int_{p_\perp^2<\xi eB}
  \frac{d^4 \tilde{p}}{(2\pi)^4} \,
  e^{-s(\tilde{p}_\parallel^2 + v_\perp^2 p_\perp^2 + m_\pi^2)} \;,
\label{eq:pion-proper}
\end{align}
where the triple meson vertex is given by
$\Gamma_{\sigma\pi\pi}(p)=-(\delta/\delta m) iD_\pi^{-1}(p)$ that we
approximate at vanishing momentum by
$\Gamma_{\sigma\pi\pi}(0)=4m/Z_\pi$.  We have used the Wick rotation
from $p$ to Euclidean $\tilde{p}$ and implemented the proper-time
regularization again with the UV cutoff $\Lambda_\pi$ that is in
principle related to the cutoff $\Lambda$ in the quark sector.  This
$p$-integration is finite and results in the following expression;
\begin{align}
 &\frac{m}{8\pi^2 v_\perp^2}\int_{1/\piLambda^2}^\infty \frac{ds}{s^2}\,
  e^{-sm_\pi^2}(1-e^{-s\xi eBv_\perp^2}) \notag\\
 &\qquad\qquad\qquad \simeq
  \frac{m}{8\pi^2}\,\xi eB\ln \frac{\Lambda_\pi^2e^{1-\gamma}}{eBv_\perp^2}
 +\mathcal{O}(m_\pi^2) \;.
\label{eq:pion-gap}
\end{align}
When the magnetic field is extremely strong, the wave-function
renormalization behaves as $Z_\pi = 8\pi^2 m^2/(eB)$ in the
leading-order of $eB$, and the velocity is
$v_\perp^2 = m^2/(eB) \ln(e^{-\gamma}\Lambda^2/m^2)$ accordingly.  The
contribution to the potential energy then becomes
\begin{align}
 V_\pi(m) &= \int^m dm' 
  \frac{m'}{8\pi^2}\,\xi eB\ln \frac{\Lambda_\pi^2e^{1-\gamma}
  \ln(e^{-\gamma}\Lambda^2/m^{\prime 2})}{m^{\prime 2}} \notag\\
 &\simeq \xi \frac{eB m^2}{16\pi^2}\ln\biggl[
  \frac{e^{2-\gamma}\Lambda_\pi^2}{m^2}
 \biggr]
  + \text{(const.)}\;,
\label{eq:pion-potential}
\end{align}
where we keep only the dominant term for small $m$ and drop negligible
terms $\propto m^2\ln(\ln m^2)$ and $m^2/\ln m^2$.  The potential
contribution from the pion loops has a singularity at $m=0$ and leads
to a divergingly positive curvature at small $m$, which favors
chiral-symmetric phase in a way opposite to the quark potential in
Eq.~(\ref{eq:V-LLL}).  Since $V_\pi(m)$ encompasses an opposite effect
to the Magnetic Catalysis, we would call this the Magnetic
Inhibition.  Interestingly the logarithmic singularity associated with
the Magnetic Inhibition is of the same strength as that with the
Magnetic Catalysis and these two effects should compete at
sufficiently strong $B$.  Our main purpose in this work is to propose
a new physical mechanism, the Magnetic Inhibition, leading to a
singularity $\sim m^2\ln m^2$, and the determination of the
singularity coefficient would require more works (and possibly depend
on details of model assumptions).

Here we would emphasize that our results are qualitatively consistent
with the latest lattice-QCD data indicating the decreasing behavior of
chiral $\Tc$ with increasing $B$.  At $T\neq 0$ the logarithmic
singularity in Eq.~\eqref{eq:V-LLL} responsible for the Magnetic
Catalysis vanishes due to the absence of the Matsubara zero-mode for
fermions and the Magnetic Catalysis is significantly
weakened~\cite{Das:1995bn,*Fukushima:2012xw}.  The Magnetic Inhibition
is, on the other hand, enhanced by the temperature effects since the
Matsubara zero-mode for bosons should be accompanied by a stronger
infrared singularity.  Therefore, the Magnetic Inhibition can soon
overcome the Magnetic Catalysis at finite temperature.  This could be
a feasible explanation for the lattice-QCD data~\cite{Bali:2011qj}.
In fact, the lattice-QCD data implies that, for a fixed value of $T$,
chiral symmetry is restored as a function of increasing $eB$.
Quantitative analyses on this phase transition induced by the Magnetic
Inhibition should deserve more investigations including lattice-QCD
simulations and chiral-model approaches.

Finally, we would stress that the present work is the first attempt
to exemplify the importance of the hadron structural change in a
strong magnetic field.  Our analysis could be extended to investigate,
for example, the possibility of the meson condensation induced by the
$B$-effect~\cite{Chernodub:2011mc}.

\begin{acknowledgments}
We thank J.~Pawlowski, A.~Rebhan, A.~Schmitt, and I.~Shovkovy for
useful discussions.  This work was supported by JSPS KAKENHI Grant
Numbers 24740169, 23340067, 24740184.
\end{acknowledgments}

\bibliography{mag}

\begin{thebibliography}{37}%
\makeatletter
\providecommand \@ifxundefined [1]{%
 \@ifx{#1\undefined}
}%
\providecommand \@ifnum [1]{%
 \ifnum #1\expandafter \@firstoftwo
 \else \expandafter \@secondoftwo
 \fi
}%
\providecommand \@ifx [1]{%
 \ifx #1\expandafter \@firstoftwo
 \else \expandafter \@secondoftwo
 \fi
}%
\providecommand \natexlab [1]{#1}%
\providecommand \enquote  [1]{``#1''}%
\providecommand \bibnamefont  [1]{#1}%
\providecommand \bibfnamefont [1]{#1}%
\providecommand \citenamefont [1]{#1}%
\providecommand \href@noop [0]{\@secondoftwo}%
\providecommand \href [0]{\begingroup \@sanitize@url \@href}%
\providecommand \@href[1]{\@@startlink{#1}\@@href}%
\providecommand \@@href[1]{\endgroup#1\@@endlink}%
\providecommand \@sanitize@url [0]{\catcode `\\12\catcode `\$12\catcode
  `\&12\catcode `\#12\catcode `\^12\catcode `\_12\catcode `\%12\relax}%
\providecommand \@@startlink[1]{}%
\providecommand \@@endlink[0]{}%
\providecommand \url  [0]{\begingroup\@sanitize@url \@url }%
\providecommand \@url [1]{\endgroup\@href {#1}{\urlprefix }}%
\providecommand \urlprefix  [0]{URL }%
\providecommand \Eprint [0]{\href }%
\providecommand \doibase [0]{http://dx.doi.org/}%
\providecommand \selectlanguage [0]{\@gobble}%
\providecommand \bibinfo  [0]{\@secondoftwo}%
\providecommand \bibfield  [0]{\@secondoftwo}%
\providecommand \translation [1]{[#1]}%
\providecommand \BibitemOpen [0]{}%
\providecommand \bibitemStop [0]{}%
\providecommand \bibitemNoStop [0]{.\EOS\space}%
\providecommand \EOS [0]{\spacefactor3000\relax}%
\providecommand \BibitemShut  [1]{\csname bibitem#1\endcsname}%
\let\auto@bib@innerbib\@empty
\bibitem [{\citenamefont {Hatsuda}\ and\ \citenamefont
  {Kunihiro}(1994)}]{Hatsuda:1994pi}%
  \BibitemOpen
  \bibfield  {author} {\bibinfo {author} {\bibfnamefont {T.}~\bibnamefont
  {Hatsuda}}\ and\ \bibinfo {author} {\bibfnamefont {T.}~\bibnamefont
  {Kunihiro}},\ }\href {\doibase 10.1016/0370-1573(94)90022-1} {\bibfield
  {journal} {\bibinfo  {journal} {Phys. Rept.}\ }\textbf {\bibinfo {volume}
  {247}},\ \bibinfo {pages} {221} (\bibinfo {year} {1994})},\ \Eprint
  {http://arxiv.org/abs/hep-ph/9401310} {arXiv:hep-ph/9401310 [hep-ph]}
  \BibitemShut {NoStop}%
\bibitem [{\citenamefont {Rischke}(2004)}]{Rischke:2003mt}%
  \BibitemOpen
  \bibfield  {author} {\bibinfo {author} {\bibfnamefont {D.~H.}\ \bibnamefont
  {Rischke}},\ }\href {\doibase 10.1016/j.ppnp.2003.09.002} {\bibfield
  {journal} {\bibinfo  {journal} {Prog. Part. Nucl. Phys.}\ }\textbf {\bibinfo
  {volume} {52}},\ \bibinfo {pages} {197} (\bibinfo {year} {2004})},\ \Eprint
  {http://arxiv.org/abs/nucl-th/0305030} {arXiv:nucl-th/0305030 [nucl-th]}
  \BibitemShut {NoStop}%
\bibitem [{\citenamefont {Fukushima}(2012)}]{Fukushima:2011jc}%
  \BibitemOpen
  \bibfield  {author} {\bibinfo {author} {\bibfnamefont {K.}~\bibnamefont
  {Fukushima}},\ }\href {\doibase 10.1088/0954-3899/39/1/013101} {\bibfield
  {journal} {\bibinfo  {journal} {J. Phys. G}\ }\textbf {\bibinfo {volume}
  {G39}},\ \bibinfo {pages} {013101} (\bibinfo {year} {2012})},\ \Eprint
  {http://arxiv.org/abs/1108.2939} {arXiv:1108.2939 [hep-ph]} \BibitemShut
  {NoStop}%
\bibitem [{\citenamefont {Alford}\ \emph {et~al.}(2008)\citenamefont {Alford},
  \citenamefont {Schmitt}, \citenamefont {Rajagopal},\ and\ \citenamefont
  {Schafer}}]{Alford:2007xm}%
  \BibitemOpen
  \bibfield  {author} {\bibinfo {author} {\bibfnamefont {M.~G.}\ \bibnamefont
  {Alford}}, \bibinfo {author} {\bibfnamefont {A.}~\bibnamefont {Schmitt}},
  \bibinfo {author} {\bibfnamefont {K.}~\bibnamefont {Rajagopal}}, \ and\
  \bibinfo {author} {\bibfnamefont {T.}~\bibnamefont {Schafer}},\ }\href
  {\doibase 10.1103/RevModPhys.80.1455} {\bibfield  {journal} {\bibinfo
  {journal} {Rev. Mod. Phys.}\ }\textbf {\bibinfo {volume} {80}},\ \bibinfo
  {pages} {1455} (\bibinfo {year} {2008})},\ \Eprint
  {http://arxiv.org/abs/0709.4635} {arXiv:0709.4635 [hep-ph]} \BibitemShut
  {NoStop}%
\bibitem [{\citenamefont {Fukushima}\ and\ \citenamefont
  {Hatsuda}(2011)}]{Fukushima:2010bq}%
  \BibitemOpen
  \bibfield  {author} {\bibinfo {author} {\bibfnamefont {K.}~\bibnamefont
  {Fukushima}}\ and\ \bibinfo {author} {\bibfnamefont {T.}~\bibnamefont
  {Hatsuda}},\ }\href {\doibase 10.1088/0034-4885/74/1/014001} {\bibfield
  {journal} {\bibinfo  {journal} {Rept. Prog. Phys.}\ }\textbf {\bibinfo
  {volume} {74}},\ \bibinfo {pages} {014001} (\bibinfo {year} {2011})},\
  \Eprint {http://arxiv.org/abs/1005.4814} {arXiv:1005.4814 [hep-ph]}
  \BibitemShut {NoStop}%
\bibitem [{\citenamefont {Kharzeev}\ \emph {et~al.}(2008)\citenamefont
  {Kharzeev}, \citenamefont {McLerran},\ and\ \citenamefont
  {Warringa}}]{Kharzeev:2007jp}%
  \BibitemOpen
  \bibfield  {author} {\bibinfo {author} {\bibfnamefont {D.~E.}\ \bibnamefont
  {Kharzeev}}, \bibinfo {author} {\bibfnamefont {L.~D.}\ \bibnamefont
  {McLerran}}, \ and\ \bibinfo {author} {\bibfnamefont {H.~J.}\ \bibnamefont
  {Warringa}},\ }\href {\doibase 10.1016/j.nuclphysa.2008.02.298} {\bibfield
  {journal} {\bibinfo  {journal} {Nucl. Phys.}\ }\textbf {\bibinfo {volume}
  {A803}},\ \bibinfo {pages} {227} (\bibinfo {year} {2008})},\ \Eprint
  {http://arxiv.org/abs/0711.0950} {arXiv:0711.0950 [hep-ph]} \BibitemShut
  {NoStop}%
\bibitem [{\citenamefont {Skokov}\ \emph {et~al.}(2009)\citenamefont {Skokov},
  \citenamefont {Illarionov},\ and\ \citenamefont {Toneev}}]{Skokov:2009qp}%
  \BibitemOpen
  \bibfield  {author} {\bibinfo {author} {\bibfnamefont {V.}~\bibnamefont
  {Skokov}}, \bibinfo {author} {\bibfnamefont {A.}~\bibnamefont {Illarionov}},
  \ and\ \bibinfo {author} {\bibfnamefont {V.}~\bibnamefont {Toneev}},\ }\href
  {\doibase 10.1142/S0217751X09047570} {\bibfield  {journal} {\bibinfo
  {journal} {Int. J. Mod. Phys.}\ }\textbf {\bibinfo {volume} {A24}},\ \bibinfo
  {pages} {5925} (\bibinfo {year} {2009})},\ \Eprint
  {http://arxiv.org/abs/0907.1396} {arXiv:0907.1396 [nucl-th]} \BibitemShut
  {NoStop}%
\bibitem [{\citenamefont {Deng}\ and\ \citenamefont
  {Huang}(2012)}]{Deng:2012pc}%
  \BibitemOpen
  \bibfield  {author} {\bibinfo {author} {\bibfnamefont {W.-T.}\ \bibnamefont
  {Deng}}\ and\ \bibinfo {author} {\bibfnamefont {X.-G.}\ \bibnamefont
  {Huang}},\ }\href {\doibase 10.1103/PhysRevC.85.044907} {\bibfield  {journal}
  {\bibinfo  {journal} {Phys. Rev.}\ }\textbf {\bibinfo {volume} {C85}},\
  \bibinfo {pages} {044907} (\bibinfo {year} {2012})},\ \Eprint
  {http://arxiv.org/abs/1201.5108} {arXiv:1201.5108 [nucl-th]} \BibitemShut
  {NoStop}%
\bibitem [{\citenamefont {D'Elia}\ \emph {et~al.}(2010)\citenamefont {D'Elia},
  \citenamefont {Mukherjee},\ and\ \citenamefont {Sanfilippo}}]{D'Elia:2010nq}%
  \BibitemOpen
  \bibfield  {author} {\bibinfo {author} {\bibfnamefont {M.}~\bibnamefont
  {D'Elia}}, \bibinfo {author} {\bibfnamefont {S.}~\bibnamefont {Mukherjee}}, \
  and\ \bibinfo {author} {\bibfnamefont {F.}~\bibnamefont {Sanfilippo}},\
  }\href {\doibase 10.1103/PhysRevD.82.051501} {\bibfield  {journal} {\bibinfo
  {journal} {Phys. Rev.}\ }\textbf {\bibinfo {volume} {D82}},\ \bibinfo {pages}
  {051501} (\bibinfo {year} {2010})},\ \Eprint {http://arxiv.org/abs/1005.5365}
  {arXiv:1005.5365 [hep-lat]} \BibitemShut {NoStop}%
\bibitem [{\citenamefont {Bali}\ \emph
  {et~al.}(2012{\natexlab{a}})\citenamefont {Bali}, \citenamefont {Bruckmann},
  \citenamefont {Endrodi}, \citenamefont {Fodor}, \citenamefont {Katz} \emph
  {et~al.}}]{Bali:2011qj}%
  \BibitemOpen
  \bibfield  {author} {\bibinfo {author} {\bibfnamefont {G.}~\bibnamefont
  {Bali}}, \bibinfo {author} {\bibfnamefont {F.}~\bibnamefont {Bruckmann}},
  \bibinfo {author} {\bibfnamefont {G.}~\bibnamefont {Endrodi}}, \bibinfo
  {author} {\bibfnamefont {Z.}~\bibnamefont {Fodor}}, \bibinfo {author}
  {\bibfnamefont {S.}~\bibnamefont {Katz}},  \emph {et~al.},\ }\href {\doibase
  10.1007/JHEP02(2012)044} {\bibfield  {journal} {\bibinfo  {journal} {JHEP}\
  }\textbf {\bibinfo {volume} {1202}},\ \bibinfo {pages} {044} (\bibinfo {year}
  {2012}{\natexlab{a}})},\ \Eprint {http://arxiv.org/abs/1111.4956}
  {arXiv:1111.4956 [hep-lat]} \BibitemShut {NoStop}%
\bibitem [{\citenamefont {Bali}\ \emph
  {et~al.}(2012{\natexlab{b}})\citenamefont {Bali}, \citenamefont {Bruckmann},
  \citenamefont {Endrodi}, \citenamefont {Fodor}, \citenamefont {Katz} \emph
  {et~al.}}]{Bali:2012zg}%
  \BibitemOpen
  \bibfield  {author} {\bibinfo {author} {\bibfnamefont {G.}~\bibnamefont
  {Bali}}, \bibinfo {author} {\bibfnamefont {F.}~\bibnamefont {Bruckmann}},
  \bibinfo {author} {\bibfnamefont {G.}~\bibnamefont {Endrodi}}, \bibinfo
  {author} {\bibfnamefont {Z.}~\bibnamefont {Fodor}}, \bibinfo {author}
  {\bibfnamefont {S.}~\bibnamefont {Katz}},  \emph {et~al.},\ }\href@noop {} {\
   (\bibinfo {year} {2012}{\natexlab{b}})},\ \Eprint
  {http://arxiv.org/abs/1206.4205} {arXiv:1206.4205 [hep-lat]} \BibitemShut
  {NoStop}%
\bibitem [{\citenamefont {Galilo}\ and\ \citenamefont
  {Nedelko}(2011)}]{Galilo:2011nh}%
  \BibitemOpen
  \bibfield  {author} {\bibinfo {author} {\bibfnamefont {B.~V.}\ \bibnamefont
  {Galilo}}\ and\ \bibinfo {author} {\bibfnamefont {S.~N.}\ \bibnamefont
  {Nedelko}},\ }\href {\doibase 10.1103/PhysRevD.84.094017} {\bibfield
  {journal} {\bibinfo  {journal} {Phys. Rev.}\ }\textbf {\bibinfo {volume}
  {D84}},\ \bibinfo {pages} {094017} (\bibinfo {year} {2011})},\ \Eprint
  {http://arxiv.org/abs/1107.4737} {arXiv:1107.4737 [hep-ph]} \BibitemShut
  {NoStop}%
\bibitem [{\citenamefont {Fraga}\ and\ \citenamefont
  {Palhares}(2012)}]{Fraga:2012fs}%
  \BibitemOpen
  \bibfield  {author} {\bibinfo {author} {\bibfnamefont {E.~S.}\ \bibnamefont
  {Fraga}}\ and\ \bibinfo {author} {\bibfnamefont {L.~F.}\ \bibnamefont
  {Palhares}},\ }\href@noop {} {\bibfield  {journal} {\bibinfo  {journal}
  {Phys. Rev.}\ }\textbf {\bibinfo {volume} {D86}},\ \bibinfo {pages} {016008}
  (\bibinfo {year} {2012})},\ \Eprint {http://arxiv.org/abs/1201.5881}
  {arXiv:1201.5881 [hep-ph]} \BibitemShut {NoStop}%
\bibitem [{\citenamefont {Fraga}\ \emph {et~al.}(2012)\citenamefont {Fraga},
  \citenamefont {Noronha},\ and\ \citenamefont {Palhares}}]{Fraga:2012ev}%
  \BibitemOpen
  \bibfield  {author} {\bibinfo {author} {\bibfnamefont {E.~S.}\ \bibnamefont
  {Fraga}}, \bibinfo {author} {\bibfnamefont {J.}~\bibnamefont {Noronha}}, \
  and\ \bibinfo {author} {\bibfnamefont {L.~F.}\ \bibnamefont {Palhares}},\
  }\href@noop {} {\  (\bibinfo {year} {2012})},\ \Eprint
  {http://arxiv.org/abs/1207.7094} {arXiv:1207.7094 [hep-ph]} \BibitemShut
  {NoStop}%
\bibitem [{\citenamefont {Schaefer}\ \emph {et~al.}(2007)\citenamefont
  {Schaefer}, \citenamefont {Pawlowski},\ and\ \citenamefont
  {Wambach}}]{Schaefer:2007pw}%
  \BibitemOpen
  \bibfield  {author} {\bibinfo {author} {\bibfnamefont {B.-J.}\ \bibnamefont
  {Schaefer}}, \bibinfo {author} {\bibfnamefont {J.~M.}\ \bibnamefont
  {Pawlowski}}, \ and\ \bibinfo {author} {\bibfnamefont {J.}~\bibnamefont
  {Wambach}},\ }\href {\doibase 10.1103/PhysRevD.76.074023} {\bibfield
  {journal} {\bibinfo  {journal} {Phys. Rev.}\ }\textbf {\bibinfo {volume}
  {D76}},\ \bibinfo {pages} {074023} (\bibinfo {year} {2007})},\ \Eprint
  {http://arxiv.org/abs/0704.3234} {arXiv:0704.3234 [hep-ph]} \BibitemShut
  {NoStop}%
\bibitem [{\citenamefont {Fukushima}(2011)}]{Fukushima:2010is}%
  \BibitemOpen
  \bibfield  {author} {\bibinfo {author} {\bibfnamefont {K.}~\bibnamefont
  {Fukushima}},\ }\href {\doibase 10.1016/j.physletb.2010.11.040} {\bibfield
  {journal} {\bibinfo  {journal} {Phys. Lett.}\ }\textbf {\bibinfo {volume}
  {B695}},\ \bibinfo {pages} {387} (\bibinfo {year} {2011})},\ \Eprint
  {http://arxiv.org/abs/1006.2596} {arXiv:1006.2596 [hep-ph]} \BibitemShut
  {NoStop}%
\bibitem [{\citenamefont {Mizher}\ \emph {et~al.}(2010)\citenamefont {Mizher},
  \citenamefont {Chernodub},\ and\ \citenamefont {Fraga}}]{Mizher:2010zb}%
  \BibitemOpen
  \bibfield  {author} {\bibinfo {author} {\bibfnamefont {A.~J.}\ \bibnamefont
  {Mizher}}, \bibinfo {author} {\bibfnamefont {M.}~\bibnamefont {Chernodub}}, \
  and\ \bibinfo {author} {\bibfnamefont {E.~S.}\ \bibnamefont {Fraga}},\ }\href
  {\doibase 10.1103/PhysRevD.82.105016} {\bibfield  {journal} {\bibinfo
  {journal} {Phys. Rev.}\ }\textbf {\bibinfo {volume} {D82}},\ \bibinfo {pages}
  {105016} (\bibinfo {year} {2010})},\ \Eprint {http://arxiv.org/abs/1004.2712}
  {arXiv:1004.2712 [hep-ph]} \BibitemShut {NoStop}%
\bibitem [{\citenamefont {Gatto}\ and\ \citenamefont
  {Ruggieri}(2011)}]{Gatto:2010pt}%
  \BibitemOpen
  \bibfield  {author} {\bibinfo {author} {\bibfnamefont {R.}~\bibnamefont
  {Gatto}}\ and\ \bibinfo {author} {\bibfnamefont {M.}~\bibnamefont
  {Ruggieri}},\ }\href {\doibase 10.1103/PhysRevD.83.034016} {\bibfield
  {journal} {\bibinfo  {journal} {Phys. Rev.}\ }\textbf {\bibinfo {volume}
  {D83}},\ \bibinfo {pages} {034016} (\bibinfo {year} {2011})},\ \Eprint
  {http://arxiv.org/abs/1012.1291} {arXiv:1012.1291 [hep-ph]} \BibitemShut
  {NoStop}%
\bibitem [{\citenamefont {Preis}\ \emph {et~al.}(2011)\citenamefont {Preis},
  \citenamefont {Rebhan},\ and\ \citenamefont {Schmitt}}]{Preis:2010cq}%
  \BibitemOpen
  \bibfield  {author} {\bibinfo {author} {\bibfnamefont {F.}~\bibnamefont
  {Preis}}, \bibinfo {author} {\bibfnamefont {A.}~\bibnamefont {Rebhan}}, \
  and\ \bibinfo {author} {\bibfnamefont {A.}~\bibnamefont {Schmitt}},\ }\href
  {\doibase 10.1007/JHEP03(2011)033} {\bibfield  {journal} {\bibinfo  {journal}
  {JHEP}\ }\textbf {\bibinfo {volume} {1103}},\ \bibinfo {pages} {033}
  (\bibinfo {year} {2011})},\ \Eprint {http://arxiv.org/abs/1012.4785}
  {arXiv:1012.4785 [hep-th]} \BibitemShut {NoStop}%
\bibitem [{\citenamefont {Preis}\ \emph {et~al.}(2012)\citenamefont {Preis},
  \citenamefont {Rebhan},\ and\ \citenamefont {Schmitt}}]{Preis:2012fh}%
  \BibitemOpen
  \bibfield  {author} {\bibinfo {author} {\bibfnamefont {F.}~\bibnamefont
  {Preis}}, \bibinfo {author} {\bibfnamefont {A.}~\bibnamefont {Rebhan}}, \
  and\ \bibinfo {author} {\bibfnamefont {A.}~\bibnamefont {Schmitt}},\
  }\href@noop {} {\  (\bibinfo {year} {2012})},\ \Eprint
  {http://arxiv.org/abs/1208.0536} {arXiv:1208.0536 [hep-ph]} \BibitemShut
  {NoStop}%
\bibitem [{\citenamefont {Klimenko}(1992{\natexlab{a}})}]{Klimenko:1990rh}%
  \BibitemOpen
  \bibfield  {author} {\bibinfo {author} {\bibfnamefont {K.}~\bibnamefont
  {Klimenko}},\ }\href {\doibase 10.1007/BF01015908, 10.1007/BF01015908}
  {\bibfield  {journal} {\bibinfo  {journal} {Theor. Math. Phys.}\ }\textbf
  {\bibinfo {volume} {89}},\ \bibinfo {pages} {1161} (\bibinfo {year}
  {1992}{\natexlab{a}})}\BibitemShut {NoStop}%
\bibitem [{\citenamefont {Klimenko}(1992{\natexlab{b}})}]{Klimenko:1992ch}%
  \BibitemOpen
  \bibfield  {author} {\bibinfo {author} {\bibfnamefont {K.}~\bibnamefont
  {Klimenko}},\ }\href {\doibase 10.1007/BF01018812, 10.1007/BF01018812}
  {\bibfield  {journal} {\bibinfo  {journal} {Theor. Math. Phys.}\ }\textbf
  {\bibinfo {volume} {90}},\ \bibinfo {pages} {1} (\bibinfo {year}
  {1992}{\natexlab{b}})}\BibitemShut {NoStop}%
\bibitem [{\citenamefont {Gusynin}\ \emph {et~al.}(1994)\citenamefont
  {Gusynin}, \citenamefont {Miransky},\ and\ \citenamefont
  {Shovkovy}}]{Gusynin:1994re}%
  \BibitemOpen
  \bibfield  {author} {\bibinfo {author} {\bibfnamefont {V.}~\bibnamefont
  {Gusynin}}, \bibinfo {author} {\bibfnamefont {V.}~\bibnamefont {Miransky}}, \
  and\ \bibinfo {author} {\bibfnamefont {I.}~\bibnamefont {Shovkovy}},\ }\href
  {\doibase 10.1103/PhysRevLett.73.3499} {\bibfield  {journal} {\bibinfo
  {journal} {Phys. Rev. Lett.}\ }\textbf {\bibinfo {volume} {73}},\ \bibinfo
  {pages} {3499} (\bibinfo {year} {1994})},\ \Eprint
  {http://arxiv.org/abs/hep-ph/9405262} {arXiv:hep-ph/9405262 [hep-ph]}
  \BibitemShut {NoStop}%
\bibitem [{\citenamefont {Gusynin}\ \emph
  {et~al.}(1995{\natexlab{a}})\citenamefont {Gusynin}, \citenamefont
  {Miransky},\ and\ \citenamefont {Shovkovy}}]{Gusynin:1994va}%
  \BibitemOpen
  \bibfield  {author} {\bibinfo {author} {\bibfnamefont {V.}~\bibnamefont
  {Gusynin}}, \bibinfo {author} {\bibfnamefont {V.}~\bibnamefont {Miransky}}, \
  and\ \bibinfo {author} {\bibfnamefont {I.}~\bibnamefont {Shovkovy}},\ }\href
  {\doibase 10.1103/PhysRevD.52.4718} {\bibfield  {journal} {\bibinfo
  {journal} {Phys. Rev.}\ }\textbf {\bibinfo {volume} {D52}},\ \bibinfo {pages}
  {4718} (\bibinfo {year} {1995}{\natexlab{a}})},\ \Eprint
  {http://arxiv.org/abs/hep-th/9407168} {arXiv:hep-th/9407168 [hep-th]}
  \BibitemShut {NoStop}%
\bibitem [{\citenamefont {Gusynin}\ \emph
  {et~al.}(1995{\natexlab{b}})\citenamefont {Gusynin}, \citenamefont
  {Miransky},\ and\ \citenamefont {Shovkovy}}]{Gusynin:1994xp}%
  \BibitemOpen
  \bibfield  {author} {\bibinfo {author} {\bibfnamefont {V.}~\bibnamefont
  {Gusynin}}, \bibinfo {author} {\bibfnamefont {V.}~\bibnamefont {Miransky}}, \
  and\ \bibinfo {author} {\bibfnamefont {I.}~\bibnamefont {Shovkovy}},\ }\href
  {\doibase 10.1016/0370-2693(95)00232-A} {\bibfield  {journal} {\bibinfo
  {journal} {Phys. Lett.}\ }\textbf {\bibinfo {volume} {B349}},\ \bibinfo
  {pages} {477} (\bibinfo {year} {1995}{\natexlab{b}})},\ \Eprint
  {http://arxiv.org/abs/hep-ph/9412257} {arXiv:hep-ph/9412257 [hep-ph]}
  \BibitemShut {NoStop}%
\bibitem [{\citenamefont {Gusynin}\ \emph {et~al.}(1996)\citenamefont
  {Gusynin}, \citenamefont {Miransky},\ and\ \citenamefont
  {Shovkovy}}]{Gusynin:1995nb}%
  \BibitemOpen
  \bibfield  {author} {\bibinfo {author} {\bibfnamefont {V.}~\bibnamefont
  {Gusynin}}, \bibinfo {author} {\bibfnamefont {V.}~\bibnamefont {Miransky}}, \
  and\ \bibinfo {author} {\bibfnamefont {I.}~\bibnamefont {Shovkovy}},\ }\href
  {\doibase 10.1016/0550-3213(96)00021-1} {\bibfield  {journal} {\bibinfo
  {journal} {Nucl. Phys.}\ }\textbf {\bibinfo {volume} {B462}},\ \bibinfo
  {pages} {249} (\bibinfo {year} {1996})},\ \Eprint
  {http://arxiv.org/abs/hep-ph/9509320} {arXiv:hep-ph/9509320 [hep-ph]}
  \BibitemShut {NoStop}%
\bibitem [{\citenamefont {Shushpanov}\ and\ \citenamefont
  {Smilga}(1997)}]{Shushpanov:1997sf}%
  \BibitemOpen
  \bibfield  {author} {\bibinfo {author} {\bibfnamefont {I.}~\bibnamefont
  {Shushpanov}}\ and\ \bibinfo {author} {\bibfnamefont {A.~V.}\ \bibnamefont
  {Smilga}},\ }\href {\doibase 10.1016/S0370-2693(97)00441-3} {\bibfield
  {journal} {\bibinfo  {journal} {Phys. Lett.}\ }\textbf {\bibinfo {volume}
  {B402}},\ \bibinfo {pages} {351} (\bibinfo {year} {1997})},\ \Eprint
  {http://arxiv.org/abs/hep-ph/9703201} {arXiv:hep-ph/9703201 [hep-ph]}
  \BibitemShut {NoStop}%
\bibitem [{\citenamefont {Gatto}\ and\ \citenamefont
  {Ruggieri}(2012)}]{Gatto:2012sp}%
  \BibitemOpen
  \bibfield  {author} {\bibinfo {author} {\bibfnamefont {R.}~\bibnamefont
  {Gatto}}\ and\ \bibinfo {author} {\bibfnamefont {M.}~\bibnamefont
  {Ruggieri}},\ }\href@noop {} {\  (\bibinfo {year} {2012})},\ \Eprint
  {http://arxiv.org/abs/1207.3190} {arXiv:1207.3190 [hep-ph]} \BibitemShut
  {NoStop}%
\bibitem [{\citenamefont {Mermin}\ and\ \citenamefont
  {Wagner}(1966)}]{Mermin:1966fe}%
  \BibitemOpen
  \bibfield  {author} {\bibinfo {author} {\bibfnamefont {N.}~\bibnamefont
  {Mermin}}\ and\ \bibinfo {author} {\bibfnamefont {H.}~\bibnamefont
  {Wagner}},\ }\href {\doibase 10.1103/PhysRevLett.17.1133} {\bibfield
  {journal} {\bibinfo  {journal} {Phys. Rev. Lett.}\ }\textbf {\bibinfo
  {volume} {17}},\ \bibinfo {pages} {1133} (\bibinfo {year}
  {1966})}\BibitemShut {NoStop}%
\bibitem [{\citenamefont {Coleman}(1973)}]{Coleman:1973ci}%
  \BibitemOpen
  \bibfield  {author} {\bibinfo {author} {\bibfnamefont {S.~R.}\ \bibnamefont
  {Coleman}},\ }\href {\doibase 10.1007/BF01646487} {\bibfield  {journal}
  {\bibinfo  {journal} {Commun. Math. Phys.}\ }\textbf {\bibinfo {volume}
  {31}},\ \bibinfo {pages} {259} (\bibinfo {year} {1973})}\BibitemShut
  {NoStop}%
\bibitem [{\citenamefont {Chodos}\ \emph {et~al.}(1990)\citenamefont {Chodos},
  \citenamefont {Everding},\ and\ \citenamefont {Owen}}]{Chodos:1990vv}%
  \BibitemOpen
  \bibfield  {author} {\bibinfo {author} {\bibfnamefont {A.}~\bibnamefont
  {Chodos}}, \bibinfo {author} {\bibfnamefont {K.}~\bibnamefont {Everding}}, \
  and\ \bibinfo {author} {\bibfnamefont {D.~A.}\ \bibnamefont {Owen}},\ }\href
  {\doibase 10.1103/PhysRevD.42.2881} {\bibfield  {journal} {\bibinfo
  {journal} {Phys. Rev.}\ }\textbf {\bibinfo {volume} {D42}},\ \bibinfo {pages}
  {2881} (\bibinfo {year} {1990})}\BibitemShut {NoStop}%
\bibitem [{\citenamefont {Agasian}\ and\ \citenamefont
  {Fedorov}(2008)}]{Agasian:2008tb}%
  \BibitemOpen
  \bibfield  {author} {\bibinfo {author} {\bibfnamefont {N.}~\bibnamefont
  {Agasian}}\ and\ \bibinfo {author} {\bibfnamefont {S.}~\bibnamefont
  {Fedorov}},\ }\href {\doibase 10.1016/j.physletb.2008.04.050} {\bibfield
  {journal} {\bibinfo  {journal} {Phys. Lett.}\ }\textbf {\bibinfo {volume}
  {B663}},\ \bibinfo {pages} {445} (\bibinfo {year} {2008})},\ \Eprint
  {http://arxiv.org/abs/0803.3156} {arXiv:0803.3156 [hep-ph]} \BibitemShut
  {NoStop}%
\bibitem [{\citenamefont {Andersen}(2012{\natexlab{a}})}]{Andersen:2012zc}%
  \BibitemOpen
  \bibfield  {author} {\bibinfo {author} {\bibfnamefont {J.~O.}\ \bibnamefont
  {Andersen}},\ }\href@noop {} {\  (\bibinfo {year} {2012}{\natexlab{a}})},\
  \Eprint {http://arxiv.org/abs/1205.6978} {arXiv:1205.6978 [hep-ph]}
  \BibitemShut {NoStop}%
\bibitem [{\citenamefont {Andersen}(2012{\natexlab{b}})}]{Andersen:2012dz}%
  \BibitemOpen
  \bibfield  {author} {\bibinfo {author} {\bibfnamefont {J.~O.}\ \bibnamefont
  {Andersen}},\ }\href@noop {} {\  (\bibinfo {year} {2012}{\natexlab{b}})},\
  \Eprint {http://arxiv.org/abs/1202.2051} {arXiv:1202.2051 [hep-ph]}
  \BibitemShut {NoStop}%
\bibitem [{\citenamefont {Das}\ and\ \citenamefont {Hott}(1996)}]{Das:1995bn}%
  \BibitemOpen
  \bibfield  {author} {\bibinfo {author} {\bibfnamefont {A.~K.}\ \bibnamefont
  {Das}}\ and\ \bibinfo {author} {\bibfnamefont {M.~B.}\ \bibnamefont {Hott}},\
  }\href {\doibase 10.1103/PhysRevD.53.2252} {\bibfield  {journal} {\bibinfo
  {journal} {Phys. Rev.}\ }\textbf {\bibinfo {volume} {D53}},\ \bibinfo {pages}
  {2252} (\bibinfo {year} {1996})},\ \Eprint
  {http://arxiv.org/abs/hep-th/9504086} {arXiv:hep-th/9504086 [hep-th]}
  \BibitemShut {NoStop}%
\bibitem [{\citenamefont {Fukushima}\ and\ \citenamefont
  {Pawlowski}(2012)}]{Fukushima:2012xw}%
  \BibitemOpen
  \bibfield  {author} {\bibinfo {author} {\bibfnamefont {K.}~\bibnamefont
  {Fukushima}}\ and\ \bibinfo {author} {\bibfnamefont {J.~M.}\ \bibnamefont
  {Pawlowski}},\ }\href@noop {} {\  (\bibinfo {year} {2012})},\ \Eprint
  {http://arxiv.org/abs/1203.4330} {arXiv:1203.4330 [hep-ph]} \BibitemShut
  {NoStop}%
\bibitem [{\citenamefont {Chernodub}(2011)}]{Chernodub:2011mc}%
  \BibitemOpen
  \bibfield  {author} {\bibinfo {author} {\bibfnamefont {M.}~\bibnamefont
  {Chernodub}},\ }\href {\doibase 10.1103/PhysRevLett.106.142003} {\bibfield
  {journal} {\bibinfo  {journal} {Phys. Rev. Lett.}\ }\textbf {\bibinfo
  {volume} {106}},\ \bibinfo {pages} {142003} (\bibinfo {year} {2011})},\
  \Eprint {http://arxiv.org/abs/1101.0117} {arXiv:1101.0117 [hep-ph]}
  \BibitemShut {NoStop}%
\end{thebibliography}%

\end{document}